\documentclass{sig-alternate}
\usepackage{url}
\begin{document}

\title{Temporal Identification of Latent Communities on Twitter}
%
%
%
%
%

\numberofauthors{6} 
%
\author{
%
%
\alignauthor
Hossein Fani\\
       \affaddr{University of New Brunswick}\\
       \email{first.last@gmail.com}
\alignauthor
Fattane Zarrinkalam\\
       \affaddr{Ferdowsi University}\\
       \email{first.last@gmail.com}
\alignauthor
Xin Zhao\\
       \affaddr{Ryerson University}\\
       \email{xin1.zhao@ryerson.ca}
\and  
\alignauthor 
Yue Feng\\
       \affaddr{Ryerson University}\\
       \email{yue.feng@ryerson.ca}
\and  
\alignauthor 
Ebrahim Bagheri\\
       \affaddr{Ryerson University}\\
       \email{bagheri@ryerson.ca}
\alignauthor
Weichang Du\\
       \affaddr{University of New Brunswick}\\
       \email{wdu@unb.ca}
}

\maketitle
\begin{abstract}
User communities in social networks are usually identified by considering explicit structural social connections between users. While such communities can reveal important information about their members such as family or friendship ties and geographical proximity, they do not necessarily succeed at pulling like-minded users that share the same interests together. In this paper, we are interested in identifying communities of users that share similar topical interests over time, regardless of whether they are explicitly connected to each other on the social network. More specifically, we tackle the problem of identifying \textit{temporal topic-based communities} from Twitter, i.e., communities of users who have similar temporal inclination towards the current topics on Twitter. We model each topic as a collection of highly correlated semantic concepts observed in tweets and identify them by clustering the time-series based representation of each concept built based on  each concept's observation frequency over time. Based on the identified topics in a given time period, we utilize multivariate time series analysis to model the contributions of each user towards the identified topics, which allows us to detect latent user communities. Through our experiments on Twitter data, we demonstrate \textit{i}) the effectiveness of our topic detection method to detect real world topics and \textit{ii})  the effectiveness of our approach compared to well-established approaches for community detection.

\end{abstract}


\section{Introduction}

Topology-based community detection methods may not be able to identify communities of users that share similar conceptual interests due to two reasons, among others: \textit{i}) There are many users on a social network that have very similar interests but are not explicitly connected to each other, e.g. through follower or followee relationships; \textit{ii}) Many of the social connections are not necessarily due to users' interests similarity but can be due to other factors such as friendship and kinship that do not necessarily point to inter-user interest similarity \cite{6}. The focus of our work in this paper is to take a temporal topic-based approach to community detection, where we focus on the temporal aspect and semantic information content of the posts to identify latent communities of users that share similar interests and similar temporal behavior but do not necessarily have explicit social interaction. 

In our work, we model the semantics of each tweet through the semantic concepts that are observed in that tweet, i.e., concepts extracted from the Tweets by a semantic annotation system. Given this representation model, we first identify and model topics  within the social network as a group of highly interrelated semantic concepts and then identify user communities based on users' interest within time with respect to the detected topics. In other words, we are interested in identifying user communities that have similar temporal inclination towards the current topics on Twitter.

For instance, when looking at Twitter data in December 2010, one can see that a set of concepts such as

\textit{New Year's Eve\footnote{\url{http://en.wikipedia.org/wiki/New\_Year's\_Eve}}, New Year's resolution, New Year's Day} and \textit{Happy New Year(Song)} form a topic to represent the New Year event. Furthermore, another set of concepts including \textit{Catherine Duchess of Cambridge, Prince William Duke of Cambridge,} and \textit{Wedding of Prince William and Catherine Middleton} form another topic to represent Prince William's Engagement. The objective of our work is to first extract these topics and then  identify implicit user communities that have similar temporal dispositions with regards to these two topics. For instance, we would like to identify the following types of communities: \textit{i}) the community of users who are interested in the topic New Year but not the other topic in the same time period. \textit{ii}) the community of users who are interested in the first topic and not the second topic `this week' from those who have the same interest pattern but in the `following week'. To this end, we propose a framework based on multivariate time series analysis to \textit{a}) identify groups of highly interrelated semantic concepts that collectively form topics on Twitter; and \textit{b}) measure inter-user similarity based on their temporal interests towards the identified topics, which is then used to identify specific user communities. The concrete contributions of our work are as follows:

\begin{enumerate}
\item Topics on Twitter are modeled as a collection of highly correlated semantic concepts. Topics are formed by clustering semantic concepts that  are represented through time series denoting the concepts' observation frequency over time. We hypothesize that those semantic concepts that have similar co-occurrence patterns through time could be correlated and can therefore collectively form a topic. 
\item Social network users are represented within a temporal multidimensional topic space whose contributions towards the identified topics over time form the user model that is utilized for identifying similar users.
\item Graph partitioning and clustering methods are used over the concept time series and multidimensional user time series to identify both the topics and the implicit user communities based on the identified topics. 

\end{enumerate}

The rest of the paper is organized as follows: The details of the proposed approach is introduced in Section 2. Section 3 is dedicated to reporting our observations from our evaluations and experiments. Section 4 reviews the related work after which we provide areas that can be improved in our work. Finally, Section 6 concludes the paper.

\section{Proposed Approach}
The main objective of our work is to identify latent user communities, within a specific time period T\footnote{ We view time period T as L consecutive time intervals.}, based on the inclination of the social network users towards the topics on Twitter. To this end, we first identify and semantically model topics within a given time period (\textit{Semantic Topic Extraction}) and then detect latent communities formed around the extracted topics (\textit{Topic-based Community Detection}). 

\subsection{Semantic Topic Extraction}
In our work, we view a \textit{topic} as a collection of temporally correlated semantic \textit{concepts} derived from external knowledge sources such as DBpedia. We utilize signal processing techniques to represent each concept's occurrence frequencies through different time intervals (\textit{concept signal}), i.e., the number of times the concept has been observed in discrete time intervals. The fundamental hypothesis behind our topic extraction method is that those concepts who have correlated concept signals could be considered to be conceptually related and can, therefore, collectively form a topic. In order to identify the correlated concepts, we exploit cross-correlation measurement, aka sliding inner-product, over pairs of concept signals. This allows us to build a \textit{concept graph} whose vertices are the observed semantic concepts and the edges denote the pairwise similarity of the source and target nodes. In order to identify topics, we apply a graph partitioning algorithm to extract coherent induced subgraphs of the concept graph. 

The core modeling paradigm of our work rests on the idea that we represent each tweet not based on the explicit observed terms in that tweet but rather through a collection of semantic concepts that have been identified in the tweet by a semantic annotator \cite{9, 10}. Therefore, in our work, each tweet is in fact a set of one or more semantic concepts that collectively denote the underlying semantics of a tweet. Now, collectively considered based on our model, users are continuously disseminating various semantic concepts through their tweets. At each point in time, a given semantic concept may be used in several different tweets posted by Twitter users; therefore, for any concept that has been mentioned at least once on Twitter, we can construct a time-domain signal that shows the number of times that the concept has been mentioned across all tweets in different time intervals of time period T, referred to as the concept signal.

\newdef{definition}{Definition}
\begin{definition}
\textbf{(Concept Signal)} A concept signal for concept $c$ is a temporally ordered set of integer values, expressed as $X_c = (x_1, x_2, …, x_L)$, from discrete observations of concept frequencies at $L$ consecutive time intervals, such that $X_c[i] = x_i = |Tweets_c^@i|$, where $Tweets_c^@i$ represents the set of all tweets posted at time interval \textit{i} that include at least one annotation referring to concept $c$.
\end{definition}

A concept signal for a semantic concept c is the occurrence frequency of c within a specific time interval. Figure 1 portrays daily concept signals of two sets of concepts referring to two real world topics: New Year Celebrations and Prince William's Engagement. We observe that concepts related to the same real world event have similar signal behavior.

\begin{figure}
\label{fig:Conceptsignals}
\centering
\includegraphics[width=0.4\textwidth]{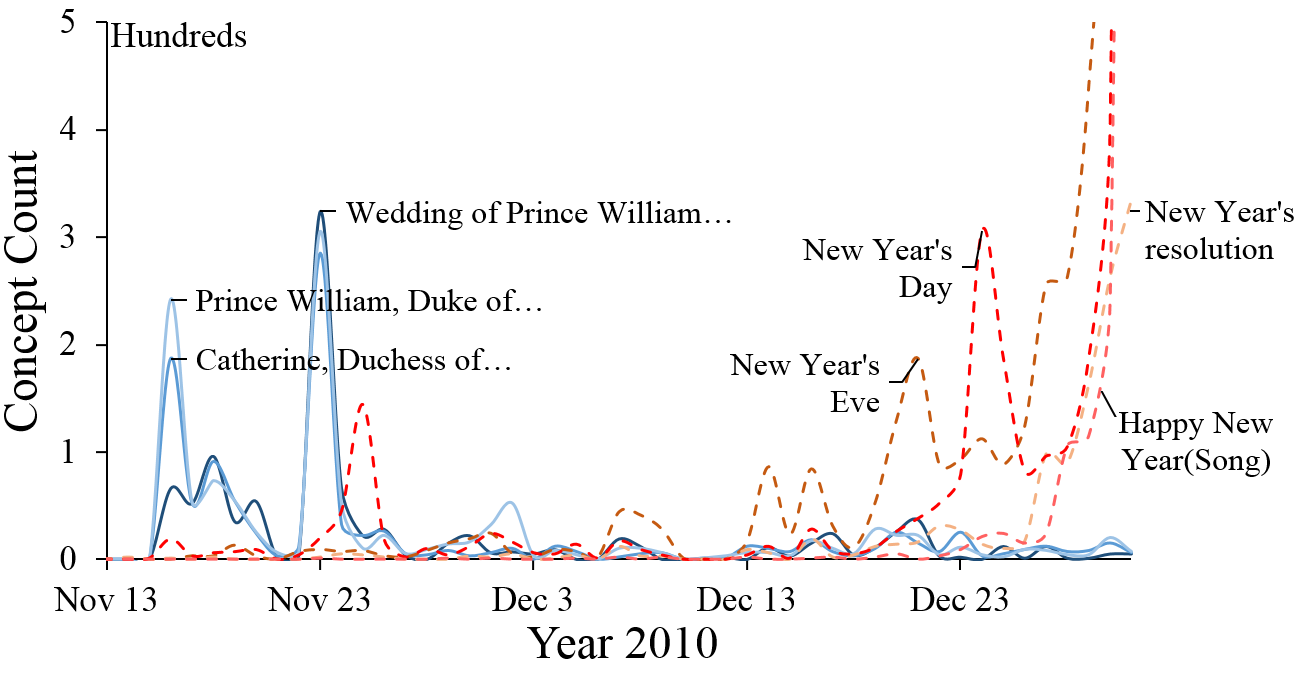}
\caption{Concept signals for two topics, New Year Celebrations and Prince William's Engagement.}
\end{figure}

Now based on Definition 1, our representation of each observed concept is through its corresponding concept signal. Therefore, we consider two concepts to be related if their respective concept signals are correlated. We infer concepts correlations by the similarity of their temporal distribution, i.e., concept signal. We calculate concepts correlations by measuring the similarity of their corresponding signals using cross-correlation.

In this work, we use cross-correlation distance score of two concepts with no time lag since we wish the correlated concepts to exhibit simultaneous similar behaviour. This is because, if the concepts are to form a topic through their correlation, they need to exhibit similar occurrence frequency patterns simultaneously. Moreover, we normalize the cross correlation to ensure the similarity between concepts pairs is in $\Re^{[0, 1]}$.

\begin{definition}
\textbf{(Concept Correlation Measure)} The concept correlation measure between two concepts $c_i$, $c_j$, denoted as $ccm(c_i,c_j)$, is defined as follows:
\begin{equation}
ccm(c_i,c_j) = \frac{X_{Ci} \star X_{Cj}}{\sqrt{\sum_{k} (X_{ci}^2[k])\times\sum_{k}(X_{cj}^2[k])}}
\end{equation}
where $X_{ci}$ and $X_{cj}$ represent concept signals for $c_i$ and $c_j$, respectively and $X_{ci} \star X_{cj}$ is a measure of the cross-correlation between two concept signals calculated as:
\begin{equation}
X_{Ci} \star X_{Cj} = \sum_{m=-\infty}^{+\infty} X_{ci}^\ast[m] X_{cj}[m]
\end{equation}
such that $ X_{ci}^\ast$ is the complex conjugate of $X_{ci}$. Since concept signals are positive, cross-correlation outputs a positive value. 
\end{definition}
Based on Definition 2, we are now able to calculate the similarity of two concepts based on the cross-correlation of their concept signals. We construct a concept graph on this basis.

\begin{definition}
\textbf{(Concept Graph)} A concept graph is a weighted undirected graph $CG= <V, E>$ where $V$ is the set of all observed semantic concepts and $E = \{ccm(c_i, c_j) | \forall  c_i, c_j \in V,  i \neq j \}$
\end{definition}

The computational time complexity of building the \textit{concept graph} is $O(\binom{|V|}{2}\times O(L))$ where $L$ is the length of the concept signal and $L \ll |V|$. Thus, the time complexity can be considered to be $O(|V|^2)$ which would not be practical for a graph of Twitter scale. We apply two filtering steps to screen out low quality concepts without impact on the topic extraction process. These filtering steps will significantly reduce the size of V and will hence make the computation of $CG.E$ practically feasible. We will introduce the details of the two filtering techniques in the evaluation section of our paper.

Now, once the concept graph is calculated, it is possible to identify highly cohesive subgraphs of CG such that they represent the topics at a given time period.

\begin{definition}
\textbf{(Topic)} Let $CG^T = (V , E)$ be a concept graph in time period $T$, we define a topic $TG= (V_{TG}, E_{TG})$ to be an induced subgraph of $CG^T$ such that $V_{TG} \subset V$  and $E_{TG}$ consists of all the edges of $CG^T$ with both end vertices in $V_{TG}$ such that $|V_{TG}|>1$ as to avoid single concept topics.
\end{definition}

In order to identify all possible topics in $CG^T$ in the form of Definition 4, we leverage a non-overlapping community detection algorithm, namely Louvain Method (LM) \cite{29} to extract cohesive subgraphs of $CG^T$ each of which would represent a topic.  Louvain is a greedy optimization method that initially finds small communities by locally maximizing modularity and consequently performs the same procedure on the new graph by considering each community extracted in the previous step as a single vertex \cite{29}. In our experiments we run the standard Louvain method (i.e., the resolution parameter is set to 1). We also ran all experiments using the VOS clustering method but since very similar results were obtained, we do not report those results in favor of space.

The outcome of the application of the community detection algorithm on the concept graph will be the identification of a set of topics that are each represented as a collection of highly correlated semantic concepts.

\subsection{Topic-based Community Detection}
After detecting \textit{Topics} from Twitter within a specific time period, our next goal is to identify latent communities of users formed on the basis of their relation to these topics. To do so, we represent the degree of contribution of a user to each topic over multiple time intervals as a vector. Therefore, given there are multiple topics in each specific time period, each user will be represented by multiple vectors, each denoting the user's contribution towards one of the topics. Collectively, this forms a multivariate signal for each user, namely the \textit{user-topic contribution signal}. Assuming there are \textit{K} topics detected, a user-topic contribution signal will be a k-variate time series. We calculate pairwise similarity between two users by computing the similarity between their corresponding user-topic contribution signals. Based on these calculated similarities, we build a weighted graph of user similarity and apply graph partitioning algorithms to detect latent communities. The contribution of a user towards a topic is defined as the frequency of the topic's concepts observed in the user's tweet set.

\begin{definition}
\textbf{(User-Topic Contribution Signal)} A \textit{user-topic contribution signal} is a k-variate time series for user $u$, denoted as $Y_u = (y_1^u,y_2^u,...,y_L^u)$ for $L$ consecutive time intervals. Then $y_t^u$ is a vector of size $K=|Topics|$, representing u's contributions at time $t$ to each of the $K$ topics: 
\begin{equation}
y_t^u[j]  =\sum_{c \in j} |Tweets_c^{@t} \cap u.tweets|.
\end{equation}
where $y_t^u[j]$ denotes the contributions of user $u$ to topic $j$ at time interval $t$. Further, $Tweets_c^{@t}$ represents the set of all tweets posted at time interval $t$ that include at least one annotation referring to concept $c$ and $u.tweets$ refers to all tweets posted by user $u$.
\end{definition}

Simply put, a \textit{user-topic contribution signal} shows, for each topic, how many times a given user has mentioned the concepts of that topic in her tweets within several consecutive time intervals, which can be visualized by a  heat map as shown in Figure 2. In this figure, the Y-axis represents the topics and the X-axis denotes the time intervals. For instance, user \url{@VegasPhotog} heavily contributed to Topic 39, represented by \textit{Hostage, Student} and \textit{Teacher} concepts (referring to an armed student who burst into a high school in Wisconsin and seized a teacher and 23 students)  on November 30 whereas user \url{@anatassara} did not react to this topic at all. All the three users mentioned in Figure 2 contributed significantly to Topic 30, which is highlighted through the \textit{New Year's Day, New Year's Resolution, New Year's Eve and Happy New Year (Song), Tornado} concepts, but with different time delays. For instance, \url{@GhorstWriter556} focused on the topic on two specific days whereas \url{@VegasPhotog} shows an increasing trend of contribution to the topic which reaches its peak on 31 Dec 2010.

\begin{figure}

\centering
\includegraphics[width=\columnwidth]{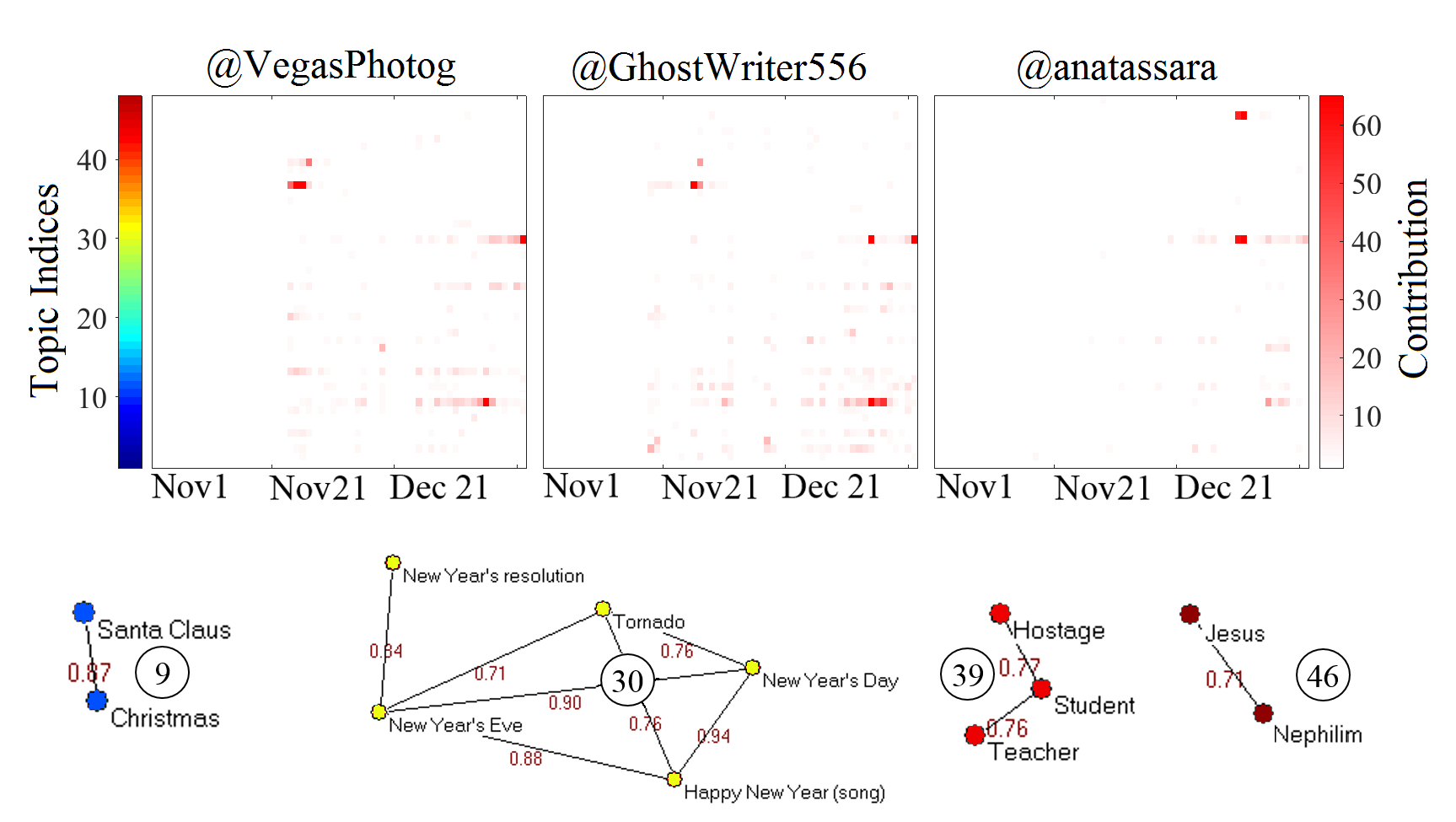}
\caption{Three user-topic contribution signals.}
\label{fig:heapmap}
\end{figure}

We believe that the behavior of the user-topic contribution signal can be considered to be a good measure for finding the similarity between two users in that it allows us to find like-minded users based on their \textit{temporally}-correlated contributions in similar topics. For instance, based on  Figure 2, the two  users, namely \url{@VegasPhotog} and \url{@GhostWriter556}, can be considered like-minded, because not only they are interested in the Topics 9, 30 and 39, but they also share similar temporal behavior with regards to these topics (e.g., both of them have contributed to Topic 9 towards the end of December). However, the third user, \url{@anatassara}, can be considered to be dissimilar from the other two because \textit{i}) she contributes to a different a topic, i.e. Topic 46, which has not received attention from the other two users; \textit{ii}) the period of time during which she reacts to Topic 9 is not completely the same as the first two users. It is noteworthy to mention that the first two users are not explicitly connected to each other on Twitter through follower/followee relationships; therefore, would have not been considered similar or placed in the same community by techniques that consider topological features of similarity.

\begin{figure}
\centering
\includegraphics[width=0.8\columnwidth]{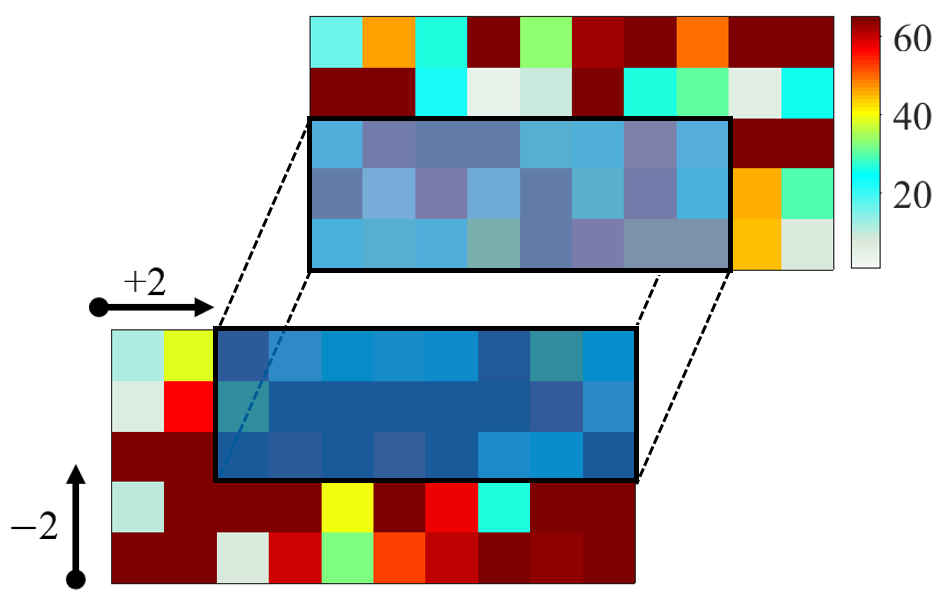}
\caption{2D cross-correlation in $XC[-2, 2]$.}
\label{fig:2Dcross}
\end{figure}

In order to compute the similarity of a pair of user-topic contribution signals, we employ the 2-dimensional variation of the cross correlation measure. Formally, the 2D cross-correlation measure of two matrices, such as $M_{[K \times L]}$ and $N_{[K \times L]}$, denoted by $ XC_{[(2K-1) \times (2L-1)]}$, is calculated as follows:
\begin{equation}
XC[i,j](M,N)= \sum_{k=0}^{K-1} \sum_{l=0}^{L-1} M[k,l] \times N^\ast[k-i,l-j]
\end{equation}
where $N^\ast$ denotes the complex conjugate of $N$. Intuitively, the 2D cross-correlation measure slides one matrix over the other and sums up the multiplications of the overlapping elements. To make it clearer, Figure \ref{fig:2Dcross} illustrates how $XC[-2, 2]$ is calculated in two $5 \times 10$ matrices. A maximum correlation occurs at $XC[0, 0]$ if the two signals are similar without any time shift. We use the normalized value of $XC[0, 0]$ in $\Re^{[0, 1]}$ when calculating user similarity distances. 

Now, given the fact that we model each user through its user-topic contribution signal, which can be represented as a $K \times L$ matrix $(Y_u)$, the similarity distance between two users can be calculated through the 2D cross-correlation of their user-topic contribution signals without a time shift.
\begin{definition}
\textbf{(User Similarity Distance)} The similarity distance of two users $u_1$ and $u_2$, denoted as $usd(u_1, u_2)$, is defined based on the 2D cross-correlation of their user-topic contribution signal as follows:
\begin{equation}
usd(u_1,u_2)= \frac {XC[0,0](Y_{u1},Y_{u2})}{(Y_{u1} \cdot Y_{u1} \times Y_{u2} \cdot Y_{u2})^{1/2}}
\end{equation}
where $Y_u$ is the user-topic contribution signal for user $u$.
\end{definition}
 
Based on Definition 6, we are now able to calculate the correlation distance between all pair of user-topic contribution multivariate signals and build a weighted \textit{user graph}.
\begin{definition}
\textbf{(User Graph)} A \textit{user graph}, $UG^T= <V, E>$, is a weighted undirected graph where $V$ is the set of all users that have tweeted in a specific time period $T$ and $E = \{usd(u_i, u_j)| \forall u_i, u_j \in V, i \neq j\}$
\end{definition}

After constructing the user graph $UG^T$ for a given time period $T$, it is possible to employ a community detection algorithms to extract clusters of users that form latent communities during time period $T$. Analogous to the topic detection phase, we utilize the Louvain method as a partitioning algorithm for this purpose.

\begin{definition}
\textbf{(Latent User Community)} Let $UG^T = (V , E)$ be a user graph in time period $T$, we define a Latent User Community $LUC = (V_{LUC} , E_{LUC})$ as an induced subgraph of $UG^T$ such that $V_{LUC} \subset V$ and $E_{LUC}$ consists of all the edges of $UG^T$ with both end vertices in $V_{LUC}$ such that $|V_{LUC}|>1$ to ensure no single user communities are permitted.
\end{definition}

The application of graph partitioning algorithms such as Louvain on $UG^T$ will produce latent user communities that consist of like-minded users, which are not necessarily topologically connected on the twitter social graph, but have contributed to the same topics with the same temporal behavior and contribution degrees.

\section{Experiments}
In this section, we describe our experiments in terms of the dataset, setup, preprocessing and comparative analytics.

\subsection{Dataset}
In our experiments, we use a Twitter dataset which is publicly available and presented by Abel et al \cite{31}. It consists of approximately 3M tweets posted by 135,731 unique users sampled between November 1 and December 31, 2010. We annotated the text of each tweet with Wikipedia concepts using the TAGME RESTful API \cite{10} which resulted in 350,731 unique concepts. As mentioned in Section 2.1, computing the edge weights and then performing community detection on a concept graph with over 300k nodes can be impractical due to the time required for computing pairwise concept correlation measurements. Therefore, we first perform a preprocessing step to identify and remove the so called \textit{stop} and \textit{white noise} concepts from our corpus. 

\subsection{Preprocessing}
In order to identify and remove uninformative concepts, we model and identify two types of concepts: i) \textit{stop concepts} and ii) \textit{white noise concepts}. Stop concepts are those that appear very frequently regardless of context or time period and therefore removing them will not perturb topic quality. On the other hand, white noise concepts are those that do not necessarily occur very often but their occurrence is randomly scattered through time without notable peaks. 

For stop concepts, given we specify them to be those concepts that are commonly observed and occur in many tweets, their temporal frequency distributions would hence follow a similar pattern to the temporal distribution of the set of all concepts in our tweets dataset. Therefore, we use the similarity of a given concept and the temporal distribution of all concepts collectively over time as an indication for being a stop concept. In other words, the more a concept signal is similar to the total number of concepts over time (AllTweets), the more likely it is to be a stop concept.

\begin{definition}
\textbf{(AllTweets Signal)} The \textit{AllTweets signal}, denoted by  $AllTweets = (x_1, x_2,..., x_L)$ is a signal such that:
$AllTweets[t] = \sum_{c \in CG.V} | Tweets_c^{@t}|$.
\end{definition}

The AllTweets signal represents the cumulative occurrence frequency of all observed semantic concepts at L consecutive time intervals. Based on Definition 9, we are now able to define stop concepts to be those concepts whose concept signal is highly correlated with the AllTweets signal.

\begin{definition}
\textbf{(Stop Concept)} A concept $sc$ is considered to be a stop concept iff:
\begin{equation}
ccm(sc, AllTweets)> \rho
\end{equation}
where $\rho$ is a threshold set to 0.9 to ensure that stop concepts are highly correlated with the AllTweets signal. 
\end{definition}

\begin{figure}
\centering
\includegraphics[width=\columnwidth]{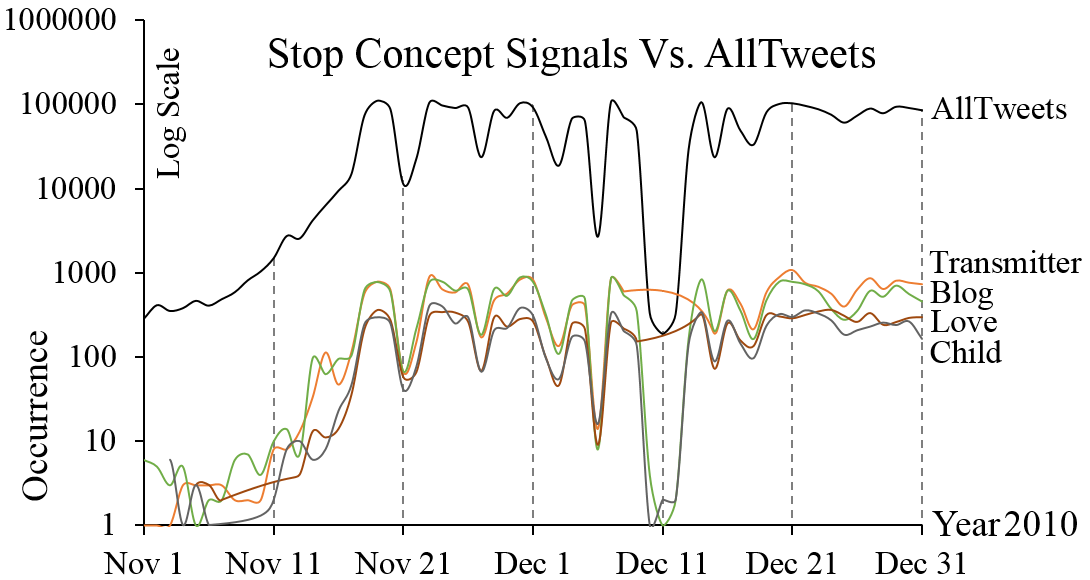}
\caption{Stop concept signals in comparison to the AllTweets signal in daily time intervals.}
\label{fig:stopConcept}
\end{figure}

Figure \ref{fig:stopConcept} depicts the behavior of some sample stop concepts in comparison to the AllTweets signal.

Now, the intuition behind \textit{white noise concepts} is that there might be many concepts that are observed within the overall Twitter timeline but their occurrence is scattered and they do not show bursty behavior at any point in time. We consider a concept \textit{wnc} to be a white noise concept if its occurrence, i.e. $X_{wnc}$, follows a normal distribution within a given time period. More formally, a white noise concept is one whose concept signal has a constant power spectral density. Provided that  the power spectral density of the concept signal is a constant value, and elements of $X_{wnc}$ have normal distribution, the concept signal would be considered to be white noise and hence can be discarded in the preprocessing phase.

In order to build such a white noise filter, we first transform all concept signals into the frequency domain to obtain their power spectral density. Then we check whether any harmonic exists in the frequency domain signal. If there is no peak detected, we conclude that this concept's power spectral density is a constant and this concept signal can be considered to be white noise. Figure 5 shows three concept signals two of which have been detected to be white noise concepts and the third is a non-filtered concept (neither stop concept nor white noise concept).

\begin{figure}
\centering
\includegraphics[width=\columnwidth]{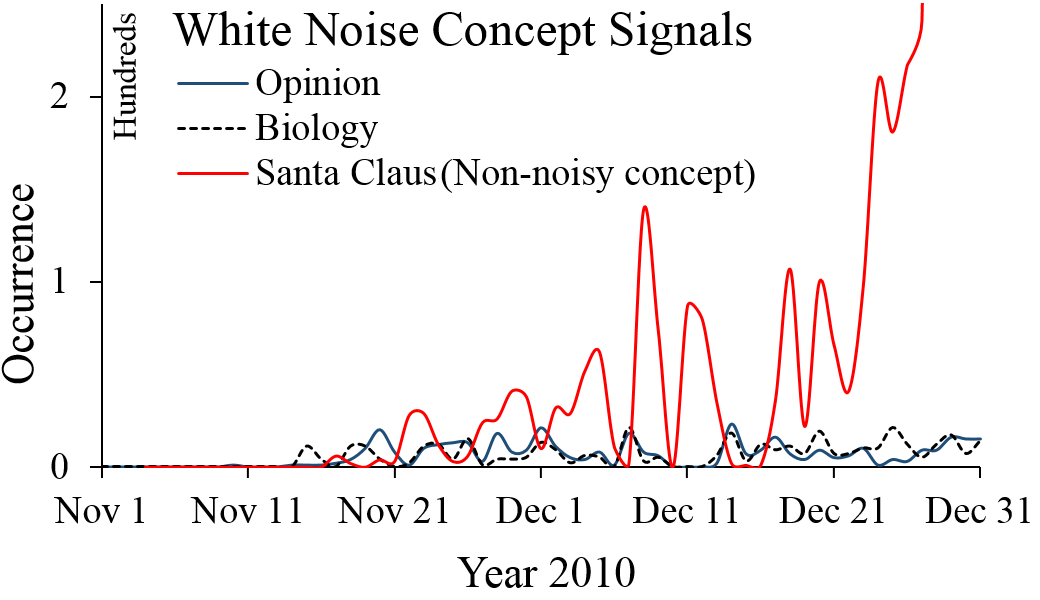}
\caption{Sample white noise in daily time intervals.}
\label{fig:whiteNoise}
\end{figure}

In our preprocessing step, the identified stop concepts and white noise concepts are removed from $CG$. These two concept filtering steps significantly reduce the size of $CG.V$ and make the computation of $CG.E$ quite practical. In our experiments the size of $CG.V$ was reduced to 782 down from over 300k initial concepts.

\subsection{Semantic Topics Evaluation}
\label{subsec:SemanticTopicsEvaluation}
There are two main parameters that can affect the performance of our topic detection approach: \textit{i}) The time interval for building concept signals based on Definition 1. In our experiments, we build the concept signals for both \textit{hourly} and \textit{daily} time intervals for the two months period of our Twitter dataset and \textit{ii}) A threshold to filter out edges in the concept graph, where the edges between those concepts that have a low correlation are removed. 
 
In order to find the best parameter setting, we have evaluated the effect of different parameters on the performance of our topic detection approach. For this purpose, we have selected \textit{Modularity} which is a common measure to calculate the quality of community detection methods, as well as the number of identified topics (i.e., \textit{Topics Count}). We believe that a good topic detection method should maximize the number of correctly identified topics while preserving a good modularity value. To this end, we have conducted an experiment through which the Louvain clustering method is applied to hourly and daily concept signals with varying thresholds for edge removal. The results of this experiment are shown in Figure 6. As shown in the figure, the hourly representation of the concept signals outperforms the daily time signals in terms of both the number of topics and modularity. Based on this analysis, we believe that the finer grained hourly representation of concept signals enables the identification of higher quality topics and therefore, in the rest of our experiments, we adopt the hourly representation of concept signals for topic detection. 

Applying the Louvain method to hourly concept signals results in higher degrees of modularity when edges with lower values are removed from the concept graph. The increase in modularity becomes noticeable when the edge weight threshold is increased to over 0.6 (the maximum value of the modularity is obtained when threshold = 0.7). Analogous to modularity, the number of topics increases by increasing the edge threshold from 0 to 0.6. However, when the value of edge threshold increases to values more than 0.6, the number of topics decreases significantly. Considering these results, we build concept signals with \textit{hourly} time intervals and set the edge threshold to 0.7. Based on these parameter settings, the modularity of our topic detection approach is more than 0.9 and the number of topics is 47. 

\begin{figure}
\centering
\includegraphics[width=\columnwidth]{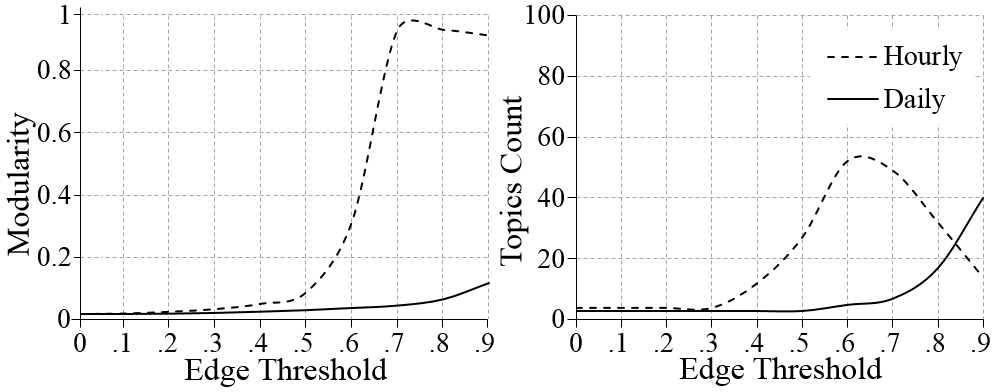}
\caption{Louvain method over hourly/daily concept signals for different edge thresholds.}
\label{fig:parameterSetting1}
\end{figure}

To evaluate the quality of the extracted topics, we manually check them, one by one, to see whether they represent real world events. The manual inspection process involved three of the authors, each of which would independently search the Web to see whether such a topic or related events to this topic in fact happened between Nov and Dec 2010. Out of the 47 topics detected, there were 9 topics which either did not represent any real world event or represented more than one event, both cases of which we considered them to be incorrect topics. From among the 9 incorrect topics, 2 represented more than one real world event (a discussion about such topics is provided in Section 5). Based on this manual analysis, the precision of our topic detection approach is $80.8\%$. We calculated the precision by measuring the number of correct topics representing exactly one real world event over the total number of topics detected. Figure 7 shows some samples of our identified topics along with the associated real world events.

\begin{figure*}
\centering
\includegraphics[width=0.8\textwidth]{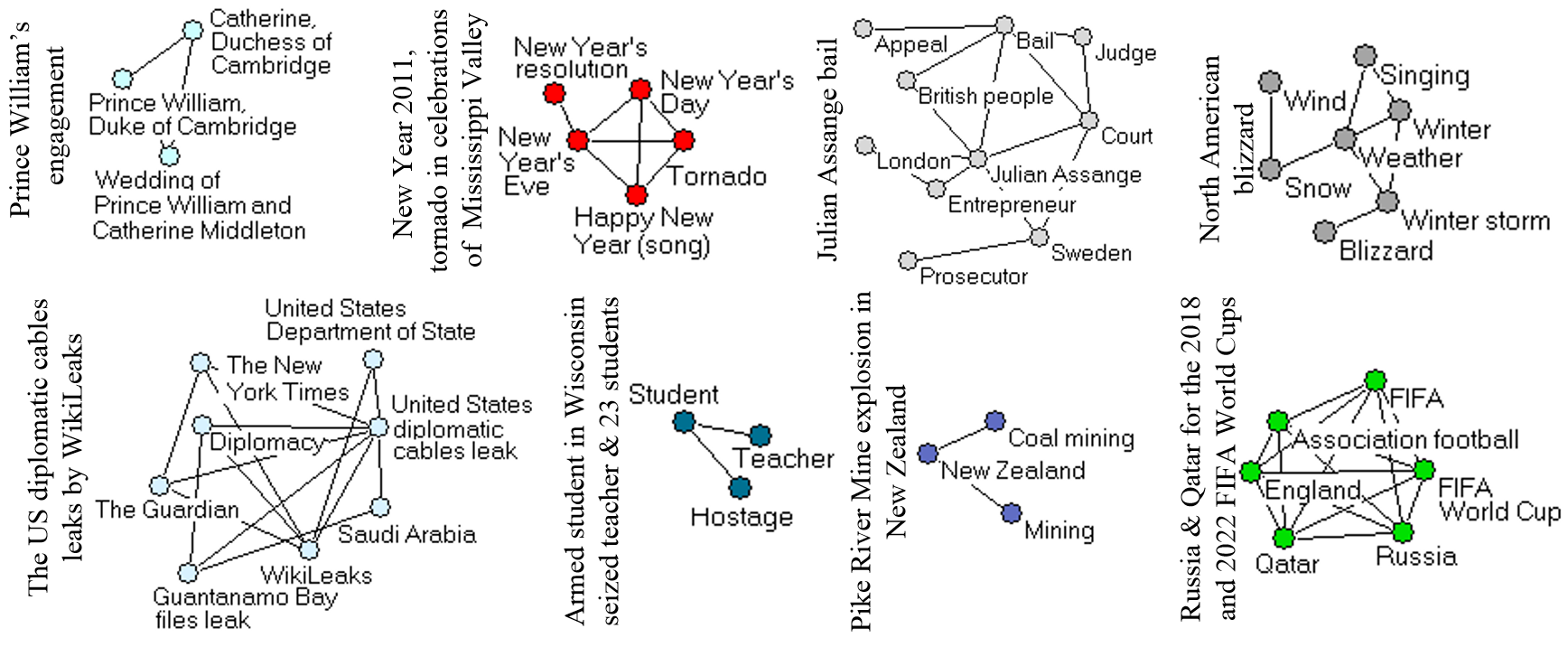}
\caption{Sample topics and the associated real world events.}
\label{fig:sampleTopics}
\end{figure*}

\subsection {User Communities Evaluation}
To evaluate the quality of the extracted communities, to the best of our knowledge and based on the related literature, there are no ground truth communities dataset that could be used to identify temporally-correlated like-minded user communities. Therefore, we evaluated the  identified communities of our approach based on the modularity metric. In the context of communities, modularity depicts whether a  community has high cohesion and low coupling. Therefore, we are interested in finding communities that have high modularity values. Furthermore, in addition to the value of modularity, we report the average community size and the number of extracted communities. We believe that a good community detection method not only has high modularity, but also produces a reasonable number of communities with a good number of users in each community.  

Similar to the topic detection process, there are two main parameters that can affect the performance of our community detection method: \textit{i}) the time interval for building user-topic contribution signal based on Definition 5. In our experiments we build these signals for both hourly and daily time intervals over the 47 identified topics from our topic detection step; and \textit{ii}) a threshold to filter out those edges that have a low weight. In order to evaluate the quality of our extracted communities and investigate how it is affected by these two parameters, we conducted experiments through which the Louvain method was applied to both hourly and daily user-topic contribution signals on different edge threshold values. The results are illustrated in Figure 8. As shown in the figure, building user-topic contribution signals based on hourly time intervals improves the value of modularity compared to daily time intervals; however, it leads to significant decrease in the average value of community size and community count for obtaining the maximum modularity value. In case of daily signals, we see that it provides better trade-off between the three measures. For example, when the edge threshold is set to 0.6, maximum modularity and maximum number of communities is obtained. Furthermore, the average communities sizes is higher in the daily time interval compared to when an hourly interval is adopted. For this reason, we believe that adopting a daily time interval could result in a better performance for community detection.

\begin{figure}
\centering
\includegraphics[width=\columnwidth]{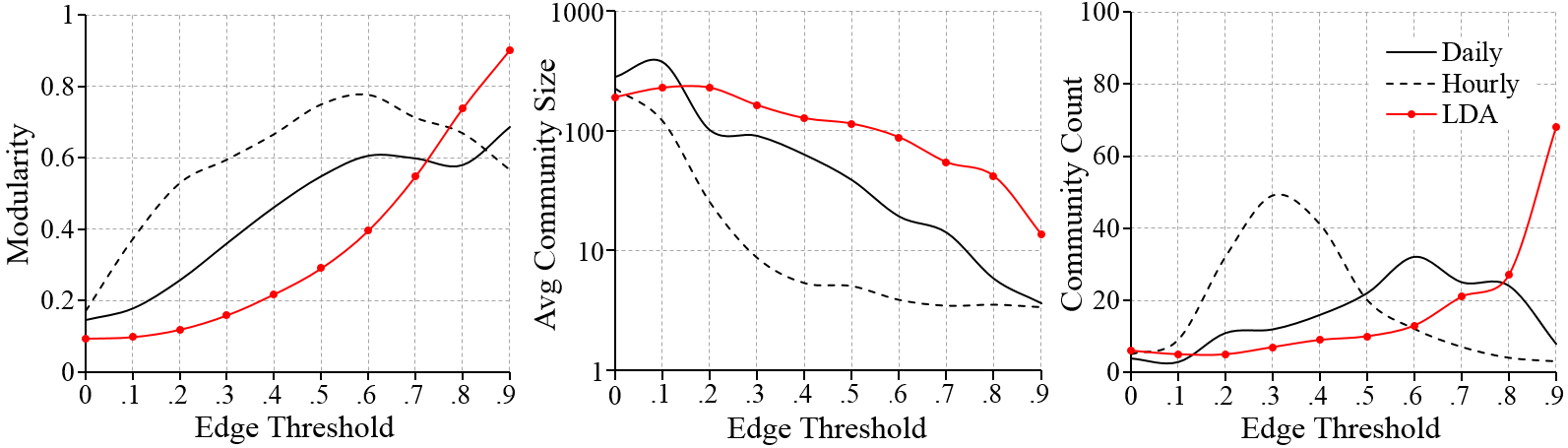}
\caption{Modularity, average community size and the number of communities over hourly and daily intervals with different edge threshold values.}
\label{fig:parameterSetting1}
\end{figure}

We further investigated the behavior of our temporal topic-based community detection by comparing our communities against both topology-based \cite{newman2004detecting} and non-temporal topic-based communities \cite{26}. For this purpose, two baseline methods are constructed as follows:

\textbf{Topology-based baseline:} We built a followership graph, which is an undirected graph whose edges are followee/follower relationships between the users. We employ the Louvain method to extract communities from the followership graph. In our experiments, we observe that the modularity of the identified communities is 0.32.

\textbf{Non-temporal topic-based baseline:} We used an LDA-based method as a non-temporal topic-based community detection. Different variations of Latent Dirichlet Allocation (LDA) have been proposed in different works to identify latent topic-based communities \cite{25, 26}. They have been widely used as a baseline for comparison \cite{1, 3, 36}. To apply LDA, we aggregated all tweets of a user as a single document and applied the MALLET implementation of LDA on the collection of such documents to discover the latent topics and probability distribution of those topics per user. We set the number of topics to 50 so that it is close to the number of topics that we identified in our approach. We then calculated the pairwise similarity of users based on the cosine similarity of their probability distributions and built a weighted user graph. Finally, the Louvain method was applied for extracting the latent user communities. The results of modularity, community count and average community size of this LDA-based community detection method for different edge threshold values are illustrated in Figure 8. 

It is important to note that Figure 8 only provides interesting insight on the behavior of each method separately and does not provides a means for comparison between the methods. This is due to the fact that the distribution of edge weights in the three methods are completely different as shown in Figure 9. For example, when using hourly signals, the value of edge weights are almost always less than 0.2. However, in case of the LDA-based method, the range of the values is much broader and includes values greater than 0.9. Low values of user similarity in our approach, especially in case of hourly time signals, is due to the fact that we calculate the similarity of two users based on their contributions to the set of topics with similar temporal behavior. In other words, in our method, two users who contribute to the same set of topics but in different time intervals would not be considered to be similar.

In order to benchmark our work, we compare the behavior of our extracted communities and the two baselines by calculating their similarity based on two popular measures for this purpose, namely Adjusted Mutual Information (AMI) and Adjusted Rand Index (ARI). The results are illustrated in Figure 10. Based on this figure, our extracted communities, whether in the case of hourly or daily time intervals, are completely different from those communities derived from topology-based method. This supports our assumption that a topology-based view to community detection on Twitter, would not necessarily be able to identify communities of users that share similar conceptual interest but do not have explicit followership relations. Our proposed approach finds latent semantic connections between users which are not explicitly connected. Therefore, in  terms of both measures, our approach is quite different from the topology-based method. However, when compared to the LDA-based method, we observe more similarity. This is because both our approach and the LDA-based method consider contributions of users towards similar topics to identify user communities; however, the outcomes of both methods are not completely the same due to the fact that our method also takes temporality of contributions into account. This is clearly observable from Figure 10 where the similarity between the daily time interval variation of our method is more similar to the LDA method compared to the hourly time interval given the impact of temporality is more severe on the hourly intervals and hence resulting in it being further away from LDA.

\begin{figure}
\centering
\includegraphics[width=\columnwidth]{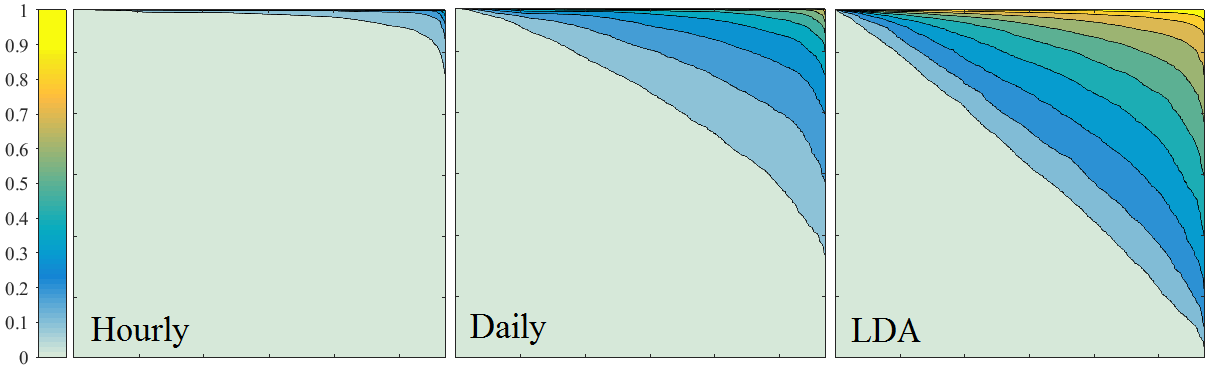}
\caption{User similarity score distributions.}
\label{fig:parameterSetting1}
\end{figure}

\begin{figure}
\centering
\includegraphics[width=\columnwidth]{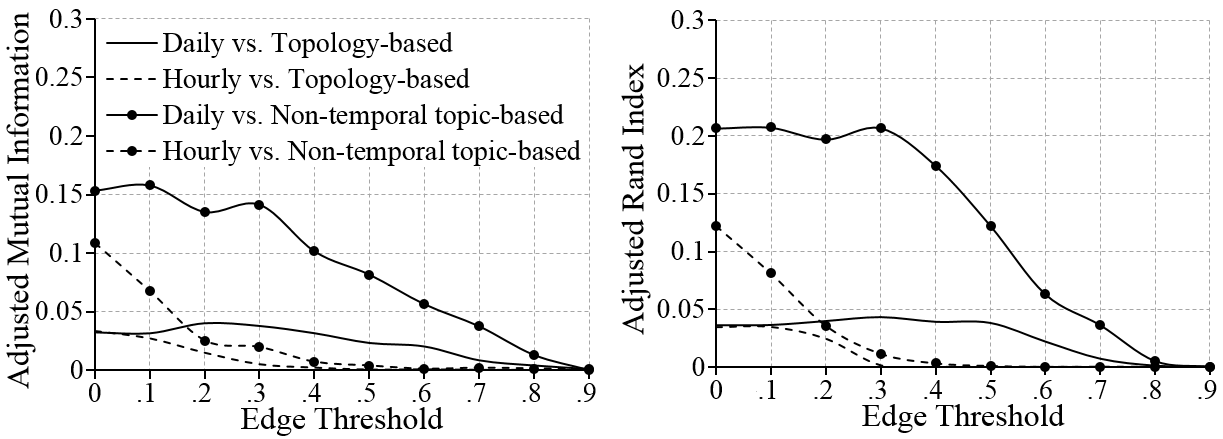}
\caption{Measures of cluster similarity between our approach and the two baselines.}
\label{fig:parameterSetting1}
\end{figure}

To highlight the significance of the temporal behavior of our approach, we depict the temporal distribution of topics over the daily topic-based communities in Figure 11. Communities are in three dimensions of \textbf{d}ay, \textbf{t}opic, and overall \textbf{c}ontribution amplitude, respectively.  As Figure 11 shows, our communities are formed not only based on different topics of interest to the users, but also based on the temporality of the user contributions. For instance, users in communities $C_1$ and $C_2$ discuss two disjoint sets of topics: \textit{Julian Assange court on Dec 15} ($t_{17}$) and \textit{Cables leak on WikiLeaks} on Nov 28 ($t_{11}$) in $C_1$ and \textit{Don't Ask, Don't Tell Repeal Act} of 2010 ($t_3$) and \textit{Thanksgiving} ($t_{37}$) in $C_2$. However, the users of communities $C_3$ and $C_4$ discuss the same topics but in different time intervals (with a week of delay).  Non-temporal approaches would merge the users of such communities ($C_3$ and $C_4$) into a single community. We believe that this an important distinguishing feature for our work. For instance, consider in the case of news recommendation, it would be unreasonable to recommend a news article to those users who discussed that topic a week ago but would make sense to recommend it to users who are currently actively pursuing it at the current point in time.

Furthermore, it is worth noting that our method is able to not only determine user communities that contribute to the same topics but also is able to distinguish between the communities that have partially overlapping topic interests. For instance, the users in $C_5$ contribute to three topics, i.e., \textit{Santa Claus} ($t_9$), \textit{New Year} ($t_{30}$) and \textit{Thanksgiving} ($t_{37}$), while users of $C_6$ also contribute to  $t_{30}$ in the same time intervals but they do not contribute to $t_9$ and $t_{30}$; therefore our method distinguishes between the users in $C_5$ and $C_6$. Similarly, this can be observed in $C_2$. In Figure 2, \url{@VegasPhotog} and \url{@GhostWriter556} end up being two members of $C_5$.

\begin{figure}
\centering
\includegraphics[width=\columnwidth]{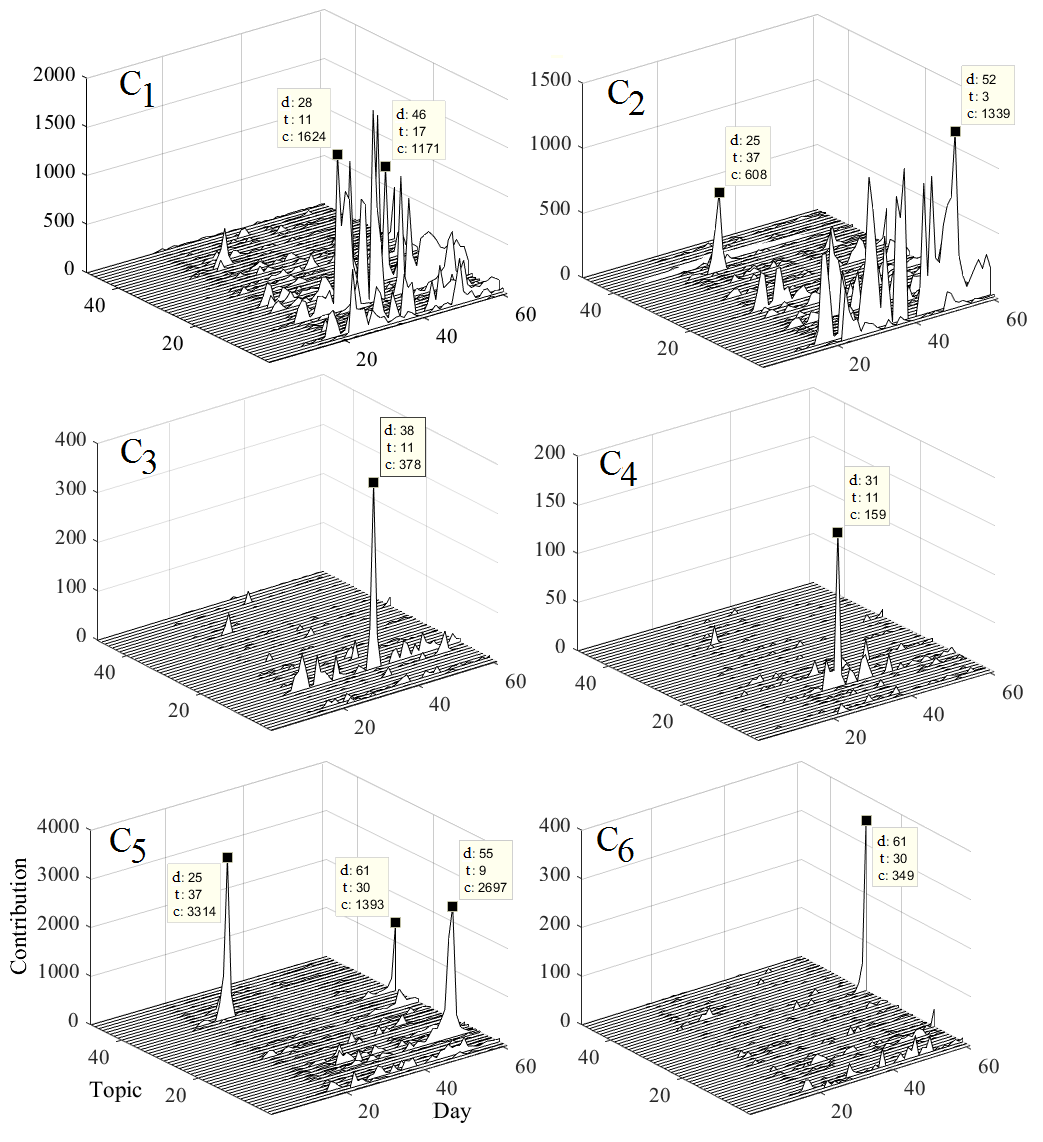}
\caption{Daily topic-based communities with edge threshold 0.6. (d,t,c denote day, topic, contribution)}
\label{fig:parameterSetting1}
\end{figure}

\section{Related Work}
\label{sec:RelatedWork}
There are two main research areas related to our work namely, \textit{topic extraction }and \textit{user community detection}.

\begin{figure}[t]
\centering
\includegraphics[width=\columnwidth]{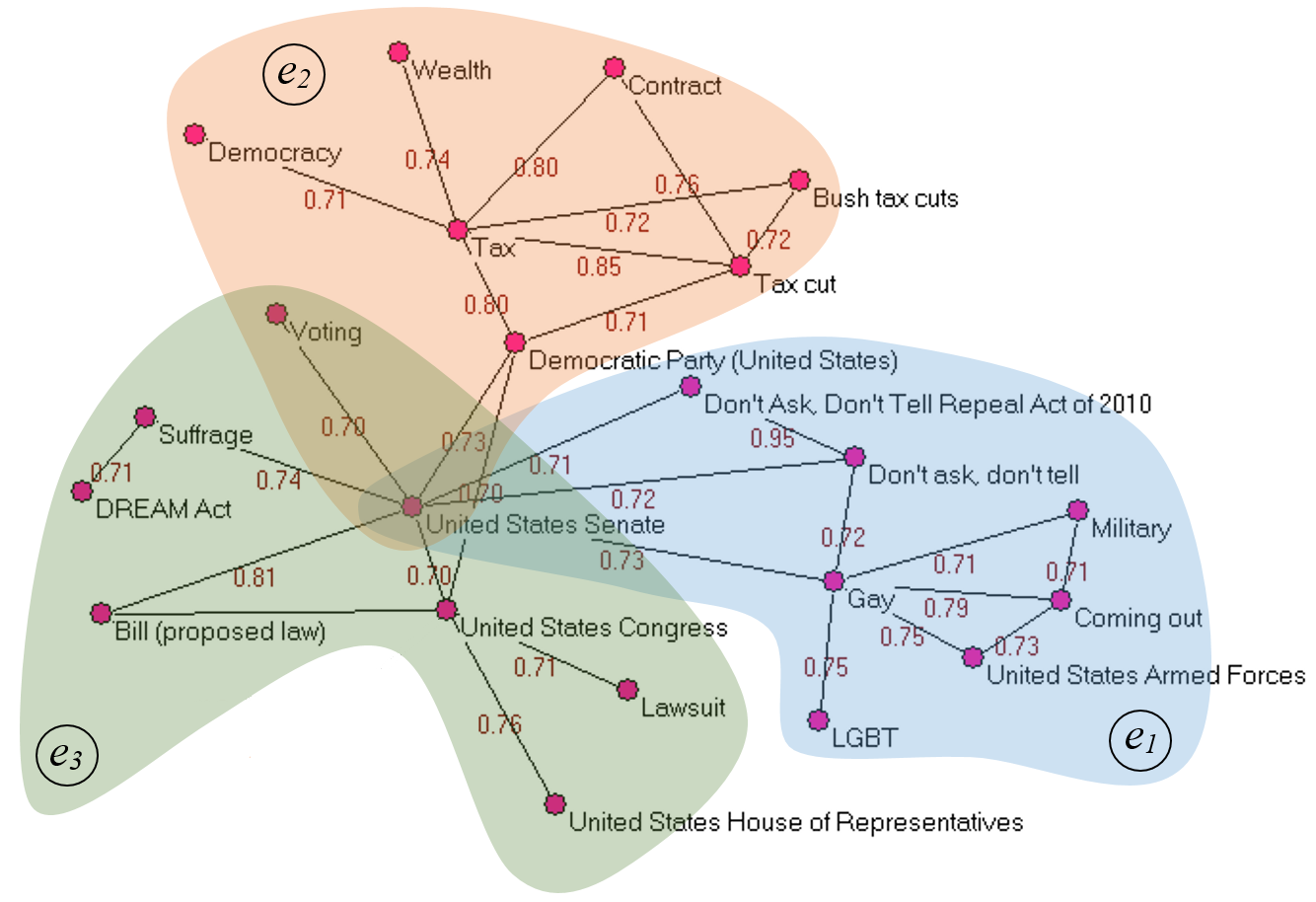}
\caption{A topic that represents three events.}
\label{fig:parameterSetting1}
\end{figure}

\subsection{Topic Extraction}
\label{subsec:TopicExtraction}
Existing topic extraction techniques can be broadly classified into three categories: document-pivot methods, topic-modeling methods and feature-pivot methods \cite{11}. Document-pivot methods extract topics by clustering documents based on the semantic distance between the documents \cite{12,13}. Topic-modeling methods (e.g. LDA) provide a probabilistic framework based on term frequencies within the documents of a given corpus. These methods form topics by extracting groups of co-occurring terms and views each document as a mixture of various topics \cite{14, 15}. Feature-pivot methods try to extract features of the topics from documents. Topics are then detected by clustering features based on their degree of semantic relatedness \cite{11}. Our proposed approach in this paper can be viewed as a feature-pivot method; therefore, we focus our review on techniques in this category.

As one of the earlier work that focused on Twitter data, Cataldi et. al. \cite{16} have constructed a co-occurrence graph of terms selected based on both the frequency of their occurrence and the importance of the users. The authors have applied a graph-based method in order to extract emerging topics. Similarly, Long et al. \cite{17} have constructed a co-occurrence graph by extracting topical words from daily posts. To extract events during a time period, they have applied a top-down hierarchical clustering algorithm over the co-occurrence graph. After detecting events in different time periods, they track changes of events in consecutive time periods and  summarize an event by finding the most relevant posts to that event. Petkos et. al. \cite{18} have argued that the algorithms that are only based on pairwise co-occurrence patterns cannot distinguish between topics which are specific to a given corpus. Therefore, they propose a soft frequent pattern mining approach to detect finer grained topics. In our own previous work \cite{19}, we have inferred fine grained users' topics of interest by viewing each topic as a conjunction of several concepts, instead of terms, and benefit from a graph clustering algorithms to extract temporally related concepts in a given time period. Further, we compute inter-concept similarity by customizing the concepts co-occurrences within a single tweet to an increased, yet semantic preserving context. 

Within the feature-pivot category, some of the work focus on time series analysis for topic detection. For instance, Weng et al. \cite{20} have used wavelet analysis to discover events in Twitter streams. First, they select bursty words by representing each word as a frequency-based signal and measure the burstiness energy of each word using autocorrelation. Then, they build a graph whose nodes are bursty words and edges are cross-correlation between each pair of bursty words and use graph-partitioning techniques to discover events. Similarly, Cordeiro \cite{21} has used wavelet analysis for event detection from Twitter. Their work constructs a wavelet signal for the hashtags, instead of words, over time by counting the hashtag mentions in each interval. Then, a continuous wavelet transformation is applied to derive a time-frequency representation of each signal. Finally, to detect an event within a given time period, peak analysis and local maxima detection techniques are employed.

\subsection{User Community Detection}
\label{subsec:UserCommunityDetection}
Existing user community detection approaches can be broadly classified into two categories \cite{5}: \textit{Topology-based} and \textit{Topic-based approaches}. The Topology-based community detection approaches represent the social network as a graph whose nodes are users and edges indicate explicit user relationships. This approach relies only on the network structure of social network graph and depends on concepts such as \textit{components} and \textit{cliques} to extract latent communities \cite{24}. On the other hand, Topic-based approaches mainly focus on information content of the users in the social network to detect latent communities. Since, the goal of our proposed approach is to detect communities formed toward the topics extracted from users' information contents, we review topic-based community detection methods in this subsection. 

Most of these works have proposed a probabilistic model to detect topic-based user   communities based on textual content or jointly with social connections \cite{2, 3, 25, 26, 27}. For example, Abdelbary et al. \cite{2} have identified users' topics of interest and extract latent communities based on the topics utilizing Gaussian Restricted Boltzmann Machine. Yin et al., \cite{36} have integrated community discovery with topic modeling in a unified generative model to detect communities of users who are coherent in both structural relationships and latent topics. In their framework, a community can be formed around multiple topics and a topic can be shared between multiple communities. Sachan et al., \cite{3} have proposed probabilistic schemes that incorporate users' posts, social connections and interaction types to discover latent user communities in social networks. In their paper, they have considered three types of interactions: a conventional tweet, a reply tweet and a re-tweet. Other authors have also proposed variations of Latent Dirichlet Allocation (LDA), for example, Author-Topic model \cite{25} and Community-User-Topic model \cite{26}, to identify latent communities. 

The above methods do not incorporate temporal aspects of users' interests and undermine the fact that users of like-minded communities would ideally show similar contribution or interest patterns for similar topics throughout time. Hu et al. \cite{37} is one of the few that consider this aspect. The authors propose a unified probabilistic generative model to extract latent communities over temporal topics and analyze topic temporal fluctuation across different communities. We also consider temporal aspects of users' interests when modeling the problem of topic-based community detection. However, We employ a 2D cross-correlation method to measure user similarity which, to our knowledge, is novel and has not been used in the latent user community detection literature. In other words, instead of\textit{ topic-distance} or \textit{degree of interest}, we define \textit{temporal user-topic contribution} and model it with multivariate time series. We extend and employ the notion of cross correlation for the users' multivariate time series in order to calculate time series similarity and partition the user base into latent user communities. As seen in \cite{37}, c.f. Figures 3 and 6, the degree of granularity of the topics identified by our approach is very fine grained and represents specific real world events, whereas the topics in \cite{37} are coarser-grained and represent higher level topics such as sports, movies and cars.

\section{Discussion}
\label{sec:Discussion}

We would like to discuss the limitations of our work and how they can be addressed in our future work: First, our topic detection algorithm is based on the cross-correlation similarity measure. Consequently, it may infer two or more completely unrelated events with disjoint set of concepts as a single coherent topic if they show similar temporal behavior. For instance as shown in Figure 12, in December 2010, three events: \textit{the Don't Ask, Don't Tell Repeal Act of 2010} ($e_1$), \textit{The Tax Relief, Unemployment Insurance Reauthorization, and Job Creation Act of 2010} ($e_2$),  and \textit{House Passes Dream Act Immigration Measures} ($e_3$), happened within the same time period and received similar contribution rates over time. For this reason, our topic detection algorithm has detected a single topic to represent these three distinct events. We plan to explore this issue in our future work by performing an additional processing step on such topics by breaking them into multiple topics. One possible solution could be that given all the concepts of a topic, we construct a concept co-occurrence graph, whose edges represent the pairwise concept co-occurrence and apply a graph clustering algorithm to discover finer-grained topics.  

The other area that requires further investigation is where a concept could potentially appear in more than one topic. For example in Figure 12, `\textit{United States Senate}' seems to be an overlapping concept in two events $e_1$ and $e_2$. However, our topic detection algorithm which uses a non-overlapping partitioning algorithm to find topics does not support overlapping concepts. It is worth noting that our topic detection algorithm which is based on concepts as elements of topics (instead of terms) is less affected by this issue. Because, in the annotation step, TagMe has already taken into account the context of tweets to annotate them with concepts defined in Wikipedia. For instance, given the tweet `\textit{\#glennbeck David Horowitz: Democratic Party Will Disappear From the Political Scene -http://bit.ly/aAvr3F}', TagMe annotates two consecutive terms \textit{democratic} and \textit{party} with a single semantic concept `\textit{democratic party(United state)}'. As a result, if we do not put `\textit{united state senate}' in the same cluster as `\textit{democratic party (United state)}' the semantics of the extracted topic would still not be hurt.

\section{Concluding Remarks}
\label{sec:Conclusions}
In this paper, we propose a framework based on multivariate time series analysis and graph partitioning methods to first identify topics on Twitter within a given time period and then detect latent communities formed with respect to the extracted topics. We model each topic as a collection of highly correlated semantic concepts and believe that the elements of a topic, i.e. concepts, show highly similar temporal behavior over the tweets space. Thus, we use concepts' signals similarity as an estimate for the semantic relatedness of pairs of concepts. On this basis, we build a weighted graph of concepts based on their similarity scores and employ graph partitioning methods to find coherent subgraph of concepts in order to represent topics. 

Given the identification of topics, we measure pairwise similarity of users based on their contributions towards the identified topics over time. We model the contribution of each user toward topics using multidimensional time series and  build a weighted user graph based on their signal similarity scores. Finally, graph partitioning methods are used to find latent communities. According to the results obtained, our topic detection method is able to effectively identify real-world events and our proposed topic-based community detection method is able to identify communities that are formed around temporally similar behavior towards topics.

\bibliographystyle{abbrv}

\begin{scriptsize}
\bibliography{WSDM2016}
\end{scriptsize}
%
\balancecolumns
\end{document}